\documentstyle[aps,preprint,epsf]{revtex} \draft
\tightenlines
\newcommand{\etal}{{\it et al.}}

\def\LP{\left(}		% left parenthesis
\def\RP{\right)}	% right parenthesis
\def\BE{\begin{equation}}
\def\EE{\end{equation}}
\def\BEA{\begin{eqnarray}}
\def\EEA{\end{eqnarray}}

\begin{document}
%@twocolumnfalse\endcsname

\title{ The static quark potential in three flavor QCD }

\author{ Claude~Bernard }
\address{
Department of Physics, Washington University, St.~Louis, MO 63130, USA
}
\author{ Tom Burch, Kostas Orginos and Doug~Toussaint }
\address{
Department of Physics, University of Arizona, Tucson, AZ 85721, USA
}
\author{ Thomas~A.~DeGrand }
\address{
Physics Department, University of Colorado, Boulder, CO 80309, USA
}
\author{ Carleton~DeTar }
\address{
Physics Department, University of Utah, Salt Lake City, UT 84112, USA
}
\author{ Steven~Gottlieb}
\address{
Department of Physics, Indiana University, Bloomington, IN 47405, USA
}
\author{ Urs~M.~Heller }
\address{
CSIT, Florida State University, Tallahassee, FL 32306-4120, USA
}
\author{ James~E.~Hetrick }
\address{
Department of Physics, University of the Pacific, Stockton, CA 95211-0197, USA
}
\author{ Bob~Sugar }
\address{
Department of Physics, University of California, Santa Barbara, CA 93106, USA
}

%Short form author list
% \author{
% Claude~Bernard$\null^a$,
% Tom Burch$\null^b$,
% Thomas~A.~DeGrand$\null^c$,
% Carleton~DeTar$\null^d$,
% Steven~Gottlieb$\null^e$,
% Urs~M.~Heller$\null^f$,
% James~E.~Hetrick$\null^g$,
% Kostas Orginos$\null^b$,
% Bob~Sugar$\null^h$
% and 
% Doug~Toussaint$\null^b$
% }
% \address{
% $\null^a$Department of Physics, Washington University, St.~Louis, MO 63130, USA,
% $\null^b$Department of Physics, University of Arizona, Tucson, AZ 85721, USA,
% $\null^c$Physics Department, University of Colorado, Boulder, CO 80309, USA
% $\null^d$Physics Department, University of Utah, Salt Lake City, UT 84112, USA,
% $\null^e$Department of Physics, Indiana University, Bloomington, IN 47405, USA,
% $\null^f$SCRI, Florida State University, Tallahassee, FL 32306-4130, USA,
% $\null^g$Department of Physics, University of the Pacific, Stockton, CA 95211-0197, USA,
% $\null^h$Department of Physics, University of California, Santa Barbara, CA 93106, USA,
% }
\date{\today}

\maketitle

\newpage
\begin{abstract}\noindent
We study the effects of dynamical quarks on the static quark potential
at distances shorter than those where string breaking is expected.
Quenched calculations and calculations with three flavors of dynamical
quarks are done on sets of lattices with the lattice spacings matched
within about one percent.  The effect of the sea quarks on the shape of
the potential is clearly visible.  We investigate the consequences of
these effects in a very crude model, namely solving Schroedinger's
equation in the resulting potential.
\end{abstract}


% OUTLINE
%  Static quark potential useful for phenomenological models.
%  Also, to set length scale
%  Effects of sea quarks:
%  1. String breaking - hard to see, need mixing with 2 meson
%  states
%  2. Slower running of coupling -> larger force at short distances
%  possibly useful to understand sea quark effects on decay constants
%  
%  Matched quenched and dynamical simulations, a=0.14
%  (matched through r_1) - explain r_1
%  Improved gauge and quark actions
%  
%  Simulation parameters - size, trajects, residuals, etc.
%  Compute potential by fixing Coulomb gauge, products of
%  time strings.
%  time distance 4-5 -- smaller showed systematic errors,
%  larger didn't help. (ESTIMATE SYSTEMATIC ERRORS?)
%  
%  Plots of matched potentials
%  plots of r_[01] sqrt(sigma)
%  
%  What difference does it make? Solve S. equation in these
%  potentials.
%  (Ref. to effect on f_pi - ask Claude?)
%  Psi(0) with string tension to set the scales.
%  Psi(0) ratios (f_pi sets scale)

\narrowtext
\section{Introduction}

One of the conceptually simplest things that can be computed in lattice QCD is the
potential between a static, infinitely heavy quark and antiquark\cite{BALI_REVIEW}.  This potential
is used in phenomenological approaches to hadron physics, and is also commonly
used in lattice studies to determine the lattice spacing, or overall energy scale.
In this work, we explore the 
differences between the potential computed in the quenched approximation and in
full QCD.

In general, there are two main effects.
The more dramatic effect of dynamical quarks is the breaking of the
string at large distances caused by shielding of the static sources by
dynamical quarks.  This string breaking is difficult to see in straightforward
potential calculations, although studies using both ``string'' type operators
and ``meson-meson'' type operators show the expected
 behavior\cite{STRINGBREAK}. We do not consider it here.
The second effect is that at
 shorter distances the dynamical quarks will change the shape of the
 potential.
Roughly speaking, this happens because the quarks reduce the QCD beta function.
This means that as we go to shorter distances, the coupling constant in full
QCD will not decrease as quickly as in quenched QCD, so the force between the
static sources will be larger in full QCD.
This effect is the subject of this paper.

This difference
is responsible for a shift of the hyperfine splitting in
quarkonia. In potential models, the splitting is proportional to the
strong coupling constant and to the square
of the wave function at the origin, which is greater
in the full theory because the coefficient of the beta function is
smaller\cite{PSI_EFFECT}.
The same effect has been
discussed as an explanation for differences in heavy-light pseudoscalar decay
constants,
$f_B$ and $f_{B_S}$, between full and quenched QCD\cite{FB_EFFECT}.
The idea is that the deeper full QCD potential at the origin will increase
 the wave function
of the light quark in a potential model, hence increasing the amplitude for
it to annihilate with the heavy quark.

When one describes the effects of the number of flavors on some
physical quantity, one is making an implicit assumption that some other
physical parameter is being held constant as the number of dynamical flavors
is varied. We are also interested in a variation of this problem ---
we would like to use a quantity determined from the heavy quark potential
as the parameter which is held constant while other physical parameters
are compared. To do this, we have to keep in mind that quenched
 QCD and full QCD
with different numbers of dynamical quark flavors are different theories.
The potentials of these theories have different shapes. It is impossible
to determine the relative length scales in the different theories by
matching the entire potential.
Instead, one must make a choice --- for example, determining the lattice
 spacing
from the string tension $\sigma$ or from Sommer's parameter $r_0$\cite{SOMMER_R0}.
If the couplings
of full and quenched simulations are tuned by matching one such parameter,
they will differ in the other parameters.

In order to isolate the effects of  dynamical quarks, we have done a series
of simulations of quenched QCD and three-flavor QCD with a range of quark
 masses,
with the gauge couplings tuned to match the lattice spacing, using one of
the possible choices of definitions of the length scale.  We used a
Symanzik improved gauge action and, in the full QCD runs, an improved
Kogut-Susskind quark action that greatly improves the scaling of physical
quantities\cite{MILC_NAIK,MILC_FAT,LEPAGE_TSUKUBA,MILC_FATTER,ILLINOIS_FAT,MILC_FATTEST,LEPAGE98,IMP_SCALING}.
The full QCD simulations were done with an optimized version of the
updating algorithm for general quark actions described in Ref.~\cite{MILC_FATTER}.
Most of the results presented here come from
a series of runs with three degenerate quarks with masses ranging from
approximately the strange quark mass to eight times this mass.  We also
include some results from shorter runs with two lighter flavors and
one quark mass fixed at the strange quark mass.
Some of the parameters of these simulations are tabulated in
 Table~\ref{RUN_TABLE}.
Lattice spacings are determined with fractional errors ranging from
$0.2$ percent for the quenched run to $0.6$
percent, and the lattice spacings are matched with a fractional
RMS spread of $0.6$ percent.
%RMS spread of UKQCD lattice spacings = 2.29%, stat. error around 1.8%

%\newpage
\renewcommand{\arraystretch}{1.5}
\begin{center}\begin{table}
\label{RUN_TABLE}
\caption{
Simulations used in this paper.  The first six columns are the sea quark mass(es), the gauge coupling, the number
of simulation time units in the run, the number of configurations for which the potential was calculated,
the size of the molecular dynamics time step, and the approximate number of conjugate gradient iterations
used in the computation of the fermion force.
The remaining columns give $r_1$, $r_0$ and $\sigma$ in units of the lattice spacing, and the
lattice spacing in fermi, using $r_1=0.36$ fm to set the scale.
The quenched runs were done with a combination of overrelaxation and heat bath steps, using
ten sweeps between samples, with each sweep consisting of four overrelaxation steps and one
heatbath step.
}
\begin{tabular}{llrrlcllll}
$am_l(/m_h)$ & $10/g^2$ & traj. & $N_{conf}$ & $\epsilon$ & CG & $r_1/a$ & $r_0/a$ & $\sigma a^2$ & $a$ (fm) \\
\hline
$\infty$ & 8.40 & na	& 38	& na	& na	& 3.739(19)	& 5.149(32)	& 0.0519(8)	& 0.0969(5)	\\
\hline
$\infty$ & 8.00 & na	& 408	& na	& na	& 2.664(6)	& 3.663(13)	& 0.1028(10)	& 0.1360(3)	\\
0.40	& 7.35	& 2004	& 330	& 0.03	& 29	& 2.661(7)	& 3.672(15)	& 0.1016(11)	& 0.1362(4)	\\
0.20	& 7.15	& 2070	& 341	& 0.03	& 46	& 2.703(7)	& 3.746(14)	& 0.0966(10)	& 0.1341(3)	\\
0.10	& 6.96	& 2064	& 339	& 0.03	& 81	& 2.687(10)	& 3.776(19)	& 0.0924(12)	& 0.1349(5)	\\
0.05	& 6.85	& 2304	& 379	& 0.02	& 127	& 2.677(11)	& 3.802(23)	& 0.0892(14)	& 0.1354(6)	\\
\hline
0.03/0.05	& 6.81	& 1164	& 190	& 0.02	& 214/143	& 2.653(15)	& 3.770(32)	& 0.0906(21) & 0.1366(8)	\\
0.02/0.05	& 6.79	& 1026	& 166	& 0.013	& 255/127	& 2.682(14)	& 3.850(32)	& 0.0852(20) & 0.1351(7)	\\
\end{tabular}
\end{table}\end{center}
\renewcommand{\arraystretch}{1.0}
%\newpage

The static quark potential including effects of sea quarks has been
calculated by the SESAM/T${\chi}$L\cite{SESAM_POT},
SCRI\cite{SCRI_POT},
UKQCD\cite{UKQCD_POT} and CPPACS\cite{CPPACS_POT} collaborations
and by Tamhankar and Gottlieb\cite{STEVES_POT}.
The SESAM calculation used the plaquette gauge action and two flavors of
Wilson quarks, with a fixed gauge coupling ($6/g^2=5.6$) for the dynamical
runs.  The SCRI group used two flavors, with both Wilson and Kogut-Susskind
 quarks,
at fixed gauge couplings.
The UKQCD calculation used the plaquette gauge action and two flavors of quarks
with the Wilson-clover action and the gauge coupling was tuned to match the
lattice spacings.  The CPPACS calculation used an improved gauge action with
two flavors of Wilson-clover quarks, varying the sea quark mass at fixed
gauge coupling.
Both the UKQCD and CP-PACS groups observed an increase in
the coefficient of the Coulomb term due to the dynamical quarks.
Because we are using three flavors of sea quarks, our results
cannot be quantitatively compared with these two flavor results.

\section{The simulations}

\begin{figure}[tbph]
\epsfxsize=4.0in
\epsfysize=4.0in
\rule{0.1in}{0.0in}\hspace{1.0in}\epsfbox[0 0 4096 4096]{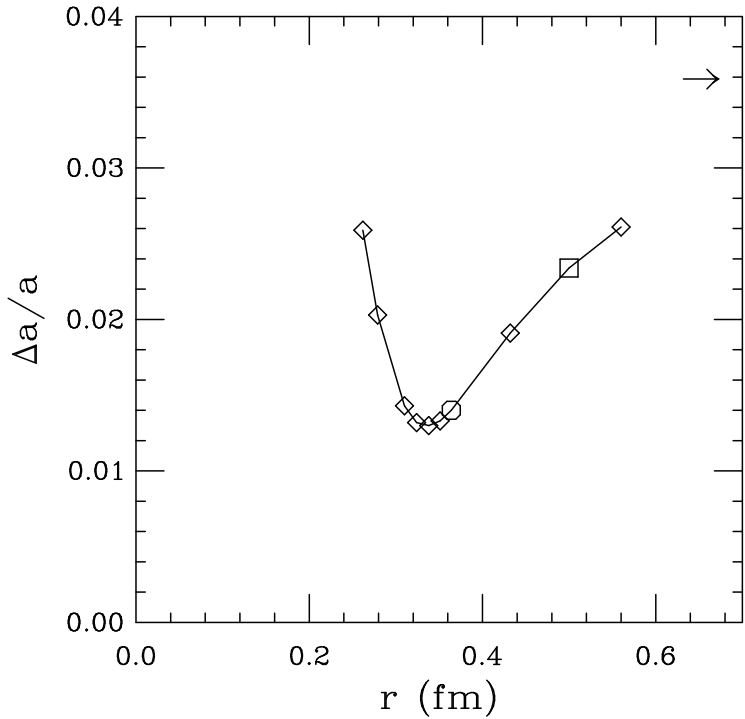}
\caption{
\label{FIX_R}
Accuracy of a lattice spacing determination from the static potential as a
function of the distance at which the force is determined.  The square is
the point where $r^2 F(r)=1.65$ and the octagon the point where
$r^2 F(r)=1$.
The arrow is the accuracy of a string tension determination.
}
\end{figure}

As mentioned above, a choice must be made when deciding how to tune the lattice
spacing.  The most common choice in recent work is to use the parameter $r_0$
defined by $r_0^2 F(r_0) = 1.65$, with $r_0 \approx 0.50 \ {\rm fm}$.  This
choice was motivated by the observation that phenomenological studies of
the potentials give consistent results for the force at this distance\cite{SOMMER_R0}.
However, as noted in the original work\cite{SOMMER_R0}, in principle $r^2 F(r)$  can be set
to any number.  In these simulations, we are faced with the challenge of tuning
the coupling with minimal effort, so it is important that the length scale we
choose be accurately measurable in a small simulation.  This motivates us to
choose a variant which we unimaginatively call $r_1$, defined by
$r_1^2 F(r_1)=1$. Figure~\ref{FIX_R} shows the fractional accuracy with which
the lattice spacing could be determined in a short simulation for different choices
of the constant in $r^2 F(r)=C$.  These are plotted as a function of the
distance at which the condition is satisfied.  The square in this plot indicates
the conventional choice $C=1.65$, and the octagon shows our choice $C=1.0$.
The arrow on the right side shows the accuracy obtained from the string tension.
The reason for the improvement in accuracy is that the potential is more accurately
determined at shorter distances.  However, at very short distances the potential is
dominated by the $-\alpha(r)/r$ part, and in this regime a length scale could
only be determined by the small dependence of $\alpha(r)$ on $r$.
Thus, the optimal choice for $C$ satisfies $r^2 F(r)=C$ at roughly
the distance where the force turns over from Coulomb-like to $\sigma r$.

\begin{figure}[p]
\epsfxsize=6.8in
\epsfysize=6.8in
\epsfbox[0 0 4096 4096]{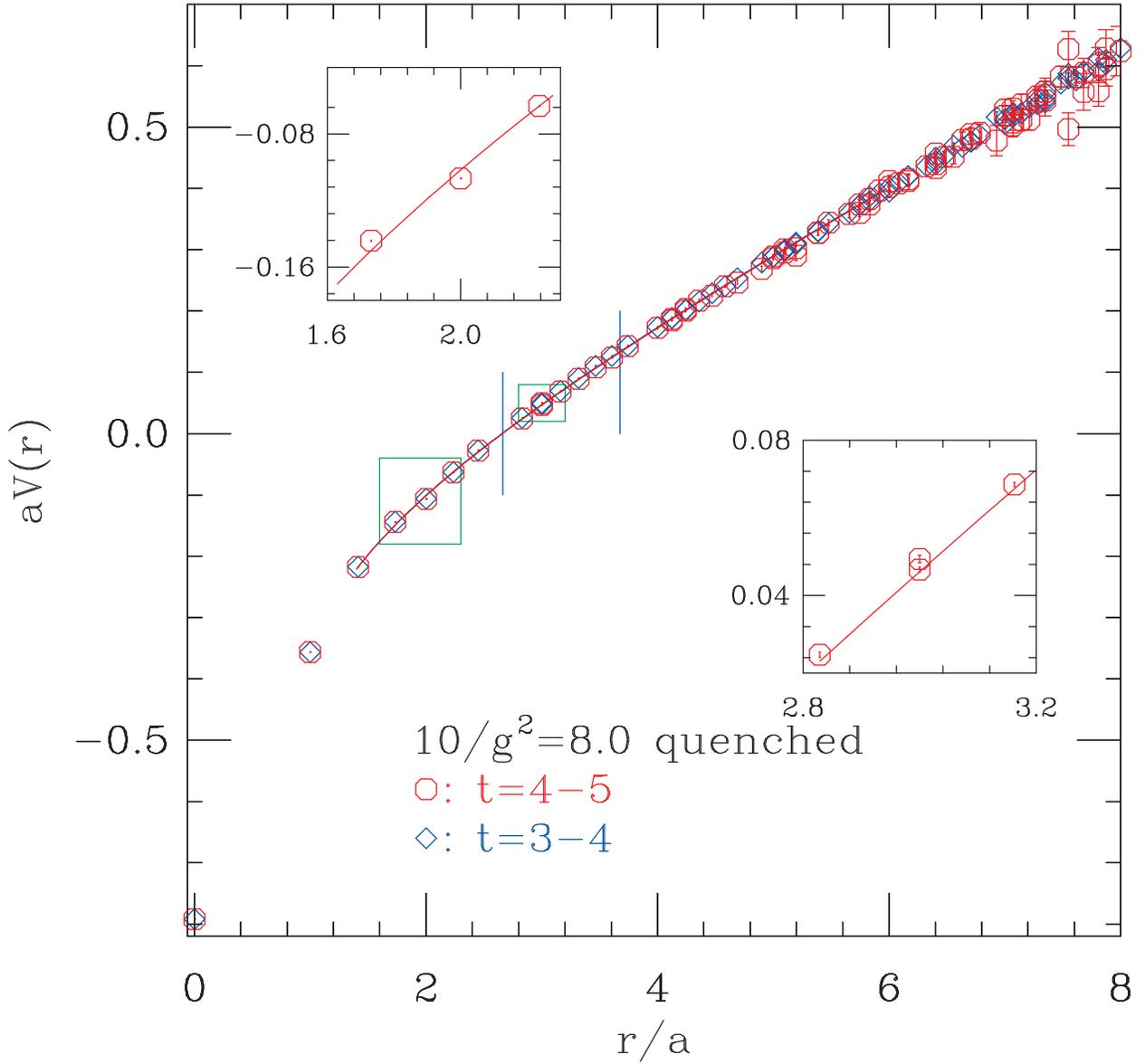}
\caption{
\label{POT_FIT_FIG}
The static quark potential in quenched QCD, with $10/g^2_{imp}=8.0$.
Octagons are from time separation four to five, and diamonds from time
separation three to four.  The solid line is the Coulomb plus linear
fit to the potential.  Insets show the remaining lattice artifacts, as
explained in the text.  A constant has been subtracted from the potential
to set $V(r_1)=0$.  The vertical lines are at $r_1$ and $r_0$.
}
\end{figure}

\begin{figure}[p]
\epsfxsize=7.0in
\epsfysize=7.0in
\epsfbox[0 0 4096 4096]{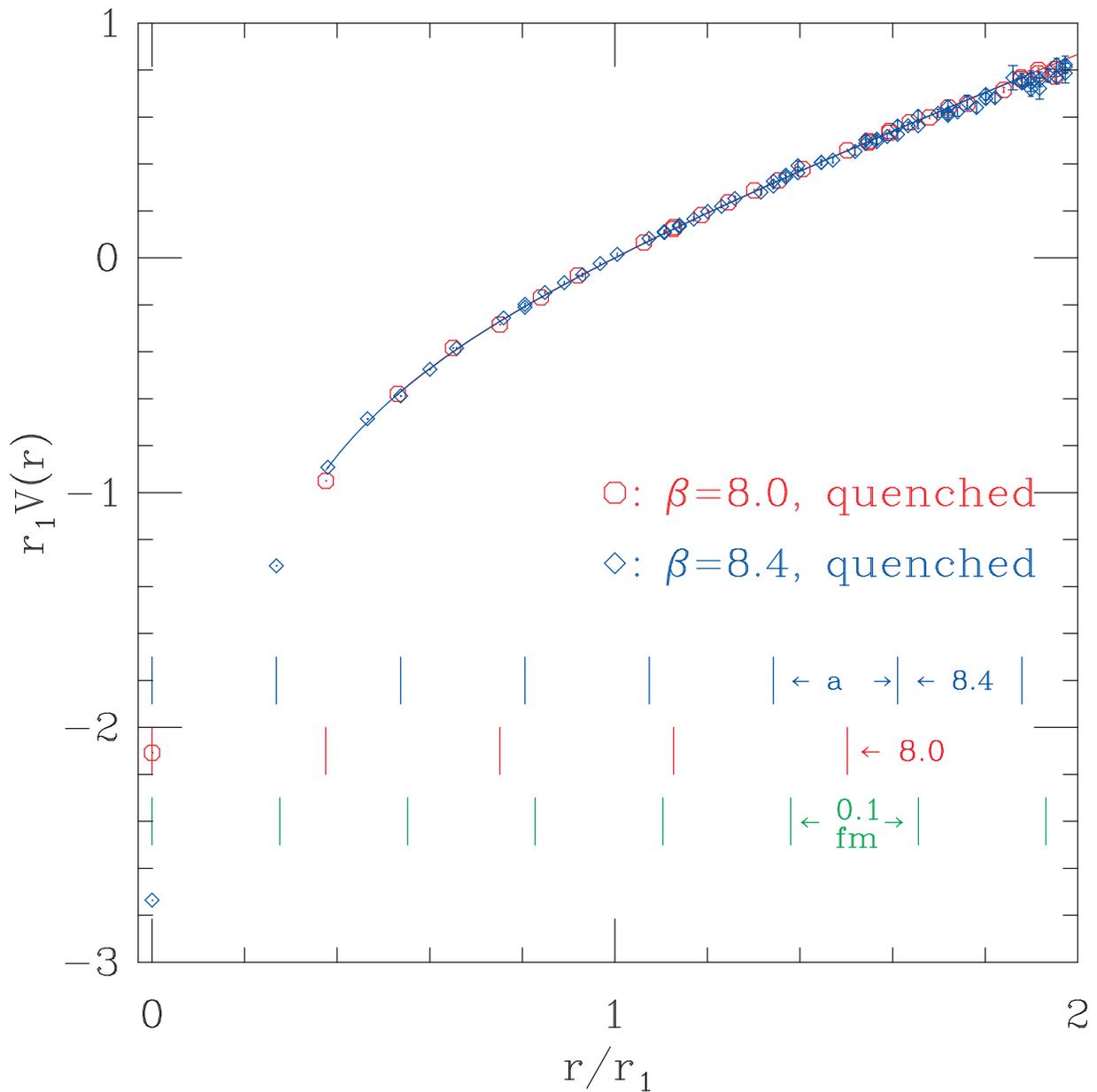}
\caption{
\label{POT_ASQ_FIG}
The short distance part of the
static quark potential in quenched QCD for two different lattice
spacings.
Octagons are from the $10/g^2=8.0$ run, and diamonds from the 
$10/g^2=8.4$ run.
The solid lines are the Coulomb plus linear
fit to the potentials, plotted in units of $r_1$.
The upper two sets of vertical lines
show the lattice spacings --- $a$, $2a$ \ldots --- for the two runs, and
the lower set is at spacings of $0.1$ fm, where the overall length
scale was set by $r_0=0.50$ fm.
}
\end{figure}

We computed the static potential by gauge fixing the lattices to the
Coulomb gauge and then just computing products of the temporal links.
This is effectively a smeared spatial string operator, smeared over
the distance scale at which the QCD dynamics correlates the links,
but we have made no further attempt to optimize our static pair creation
operator.  The potential was determined from the ratios of ``loops''
at time separations four and five
(five and six for the $a\approx 0.10$ quenched run).
Ratios of distances three and
four gave results which in some cases differed from the ratios at
times 4-5 by more than the statistical error, while inclusion of
larger distances in a fit did not appreciably improve the statistical
errors.  After the potential was computed at all spatial separations,
separations related by lattice symmetries were combined.
Then the potential was fit to the form $V(r) = C - \alpha/r + \sigma r$.
This form gave good results for separations from $(1,1,0)$ to the largest
values for which we could compute the potential.  Our quoted results are
obtained to fits from all points at spatial distances from $\sqrt{2} a$ to $6 a$,
with the exception of the $a \approx 0.10$ fm run, where distances up to 
$7a$ were used.
$r_0$ and $r_1$ were
derived from $\alpha$ and $\sigma$.  Such a simple approach to fitting
the potential is only possible because the improved action suppresses most
of the lattice artifacts.  Figure~\ref{POT_FIT_FIG} shows the potential
for the quenched run, where our statistics are largest, together with
the fit that was found.  A remaining lattice artifact is visible at distance
two (separation $(2,0,0)$), where the potential is visibly below the smooth
curve.  An elegant test for lattice artifacts is to compare the two points
at distance three, at separations $(3,0,0)$ and $(2,2,1)$, which are shown
in an inset in Fig.~\ref{POT_FIT_FIG}.  For the quenched potential there
is a small discrepancy here, with the ``on-axis'' point lower.
For most of our full QCD runs this discrepancy is smaller than the statistical
errors.
For the quenched case we have done a shorter simulation at a smaller lattice
spacing, $a \approx 0.10$ fm to check for dependence on the lattice
spacing.  This run was done on $28^3\times 96$ lattices at $10/g^2=8.4$.
Figure~\ref{POT_ASQ_FIG} shows the short distance part of the potential
for the quenched theory at the two lattice spacings, showing excellent
agreement.
Since the main point of this work is to isolate the effects of
dynamical quarks, we have not
included a systematic error from the remaining lattice artifacts in our 
quoted error estimates.  First, we think that this error is small.  But
more importantly, we are comparing quenched and full QCD lattices at matched
lattice spacings, and these systematic errors will largely cancel in the
differences of the two.  (All of our runs show essentially the same
discrepancy at distance two, for example.)

As detailed in Table~\ref{RUN_TABLE}, we were able to choose values of the
gauge coupling to match $r_1$ in the different runs with a fractional
RMS spread of $0.6$ percent, with the final lattice spacings known with
accuracies ranging from $0.2$ percent for the quenched run to $0.6$ percent
for the lowest mass run.
%RMS spread of UKQCD lattice spacings = 2.29%, stat. error around 1.8%
%RMS spread of MILC  lattice spacings = 0.61%, stat. error around 0.5%

Our procedure for determining the lattice spacing differs from that used
in Ref.~\cite{SOMMER_R0} (see also \cite{GUAGNELLI}),
where an interpolation is done using values of the potential along one axis,
using a tree level calculation of the potential to correct for lattice artifacts.
We have compared the results of the different methods using our large $10/g^2=8.0$
quenched sample.  In Table~\ref{METHODS} we tabulate $r_0/a$ and $r_1/a$ calculated
using our fitting procedure and by the procedure of Ref.~\cite{SOMMER_R0}
applied along the lattice directions $(1,0,0)$, $(1,1,0)$ and $(1,1,1)$.
The results are consistent, with $r_1/a$ better determined than $r_0/a$
for both methods.  The fitting method produces smaller statistical errors simply
because it makes use of the potential at all possible separations.

All of the errors quoted in this paper are statistical errors obtained with
the jackknife method.

\begin{center}\begin{table}
\label{METHODS}
\caption{
Accuracy and consistency
of the lattice spacing determination from fitting all points
and from the method of Ref.~\protect\cite{SOMMER_R0} applied along three
different axes.  For the method of Ref.~\protect\cite{SOMMER_R0} the
second column shows the direction along which interpolation
was done.
}
\begin{tabular}{llll}
Quantity & Method & Value & Fractional error \\
%$r_0/a$     & FIT        & 3.6633(125)     & 0.0034 \\
%$r_0/a$     & $(1,0,0)$  & 3.6660(208)     & 0.0057 \\
%$r_0/a$     & $(1,1,0)$  & 3.6906(272)     & 0.0074 \\
%$r_0/a$     & $(1,1,1)$  & 3.8760(932)     & 0.0240 \\
$r_0/a$     & FIT        & 3.663(13)     & 0.0034 \\
$r_0/a$     & $(1,0,0)$  & 3.666(21)     & 0.0057 \\
$r_0/a$     & $(1,1,0)$  & 3.691(27)     & 0.0074 \\
$r_0/a$     & $(1,1,1)$  & 3.876(93)     & 0.0240 \\
\vspace{0.01in}\\
%$r_1/a$     & FIT        & 2.6644(60)      & 0.0023 \\
%$r_1/a$     & $(1,0,0)$  & 2.6503(88)      & 0.0033 \\
%$r_1/a$     & $(1,1,0)$  & 2.6609(135)     & 0.0051 \\
%$r_1/a$     & $(1,1,1)$  & 2.6921(186)     & 0.0069 \\
$r_1/a$     & FIT        & 2.664(6)      & 0.0023 \\
$r_1/a$     & $(1,0,0)$  & 2.650(9)      & 0.0033 \\
$r_1/a$     & $(1,1,0)$  & 2.661(14)     & 0.0051 \\
$r_1/a$     & $(1,1,1)$  & 2.692(19)     & 0.0069 \\
\end{tabular}\end{table}\end{center}

\widetext
\newpage
\newpage
\narrowtext

\widetext
\newpage
\newpage
\narrowtext

\section{Effects of sea quarks}

\begin{figure}[tbp]
\epsfxsize=7.0in
\epsfysize=7.0in
\epsfbox[0 0 4096 4096]{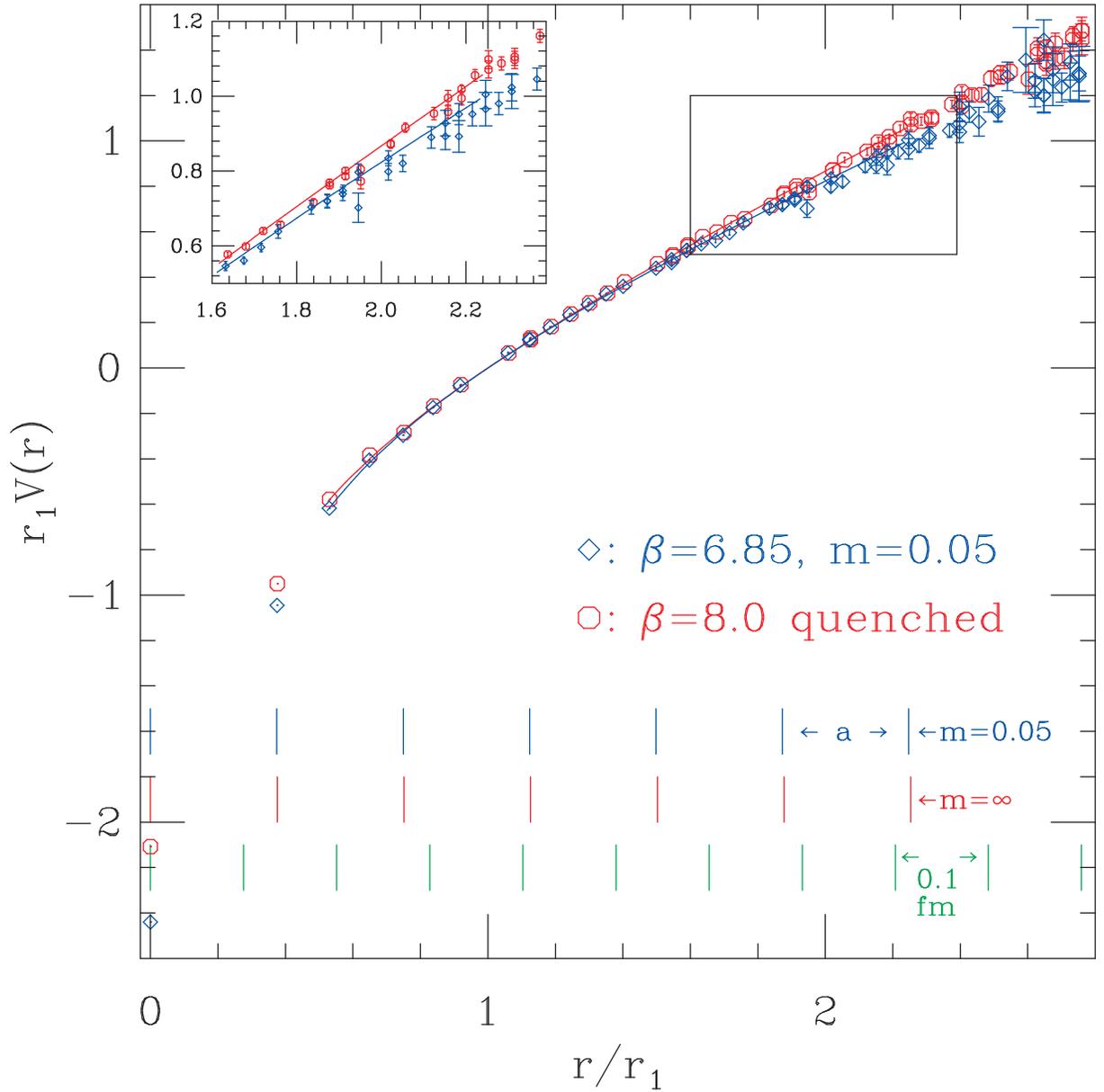}
\caption{
\label{CMP_PLOT_FIG}
The static quark potential for quenched (octagons) and three flavor (diamonds) QCD, in
units of $r_1$.  The solid lines are fits to the Coulomb plus constant plus linear
form.  The lattice spacing was matched using $r_1$ as described in the text.
As in Fig.~\protect\ref{POT_ASQ_FIG}, the upper two rulers show the lattice
spacing in the two runs, and the lower one shows units of $0.1$ fm.
The inset expands the area shown by the box.
}
\end{figure}

In Fig.~\ref{CMP_PLOT_FIG}, we plot the quenched potential and the potential
with three degenerate flavors of quarks at $am_q=0.05$.  This is approximately
the mass of the strange quark.
Both the distance scale and the potential are
plotted in units of $r_1$, and a constant has been subtracted from the
potential so that it is zero at $r_1$.  Since $r_1$ was determined from this
potential, this results in matching the slope of the potentials at $r=r_1$,
so the fits are tangent at this point.
Away from the point where the force
is matched, the potentials have different slopes, as expected.  
This graph contains three ``rulers''.
The upper sets of vertical lines
show the lattice spacings for the dynamical and quenched runs, and
the lower set is at spacings of $0.1$ fm, where the overall length
scale was set by $r_0=0.50$ fm in the quenched run.

\widetext
\newpage
\newpage
\narrowtext

\begin{figure}[tbp]
\epsfxsize=4.0in
\epsfysize=4.0in
\rule{0.1in}{0.0in}\hspace{1.0in}\epsfbox[0 0 4096 4096]{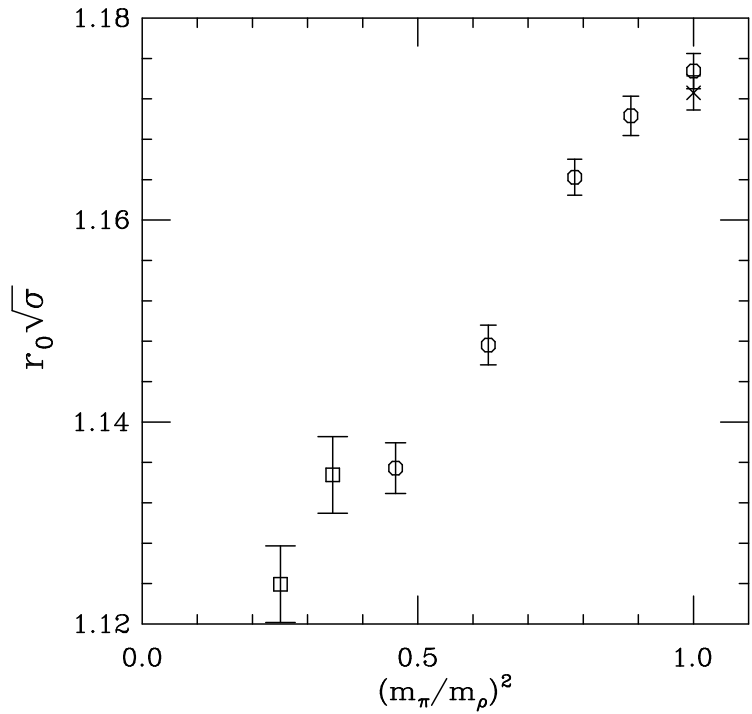}
\caption{
\label{R0_SIGMA_FIG}
Effects of dynamical quarks on the shape of the potential.  Here we plot
$r_0 \sqrt{\sigma}$ as a function of the quark mass.  The quenched points
are at the right, with the octagon coming from the $10/g^2=8.0$ run and
the cross from the $10/g^2=8.4$ run.  The remaining octagons are
full QCD runs with three degenerate flavors, and the squares are full
QCD runs with two light flavors and one heavy.
}
\end{figure}

\begin{figure}[tbp]
\epsfxsize=4.0in
\epsfysize=4.0in
\rule{0.1in}{0.0in}\hspace{1.0in}\epsfbox[0 0 4096 4096]{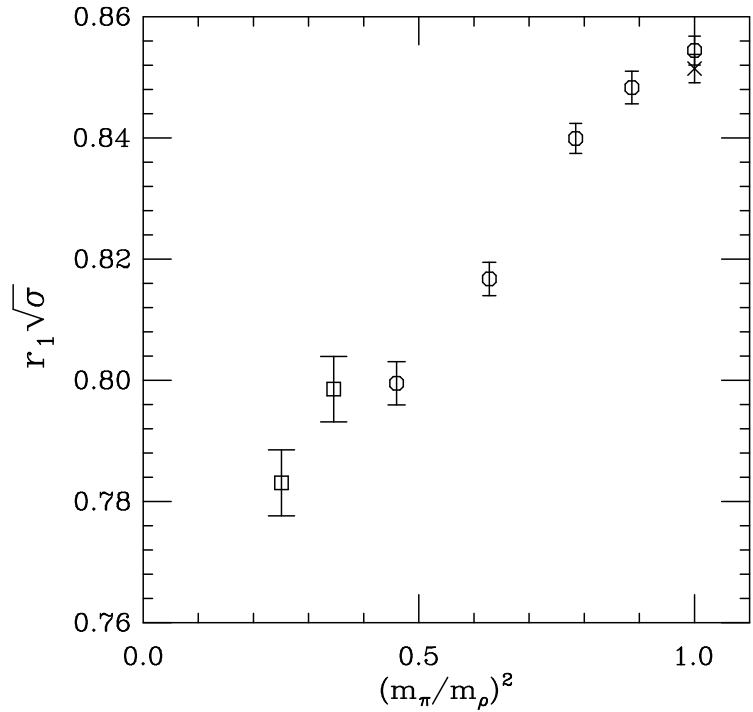}
\caption{
\label{R1_SIGMA_FIG}
The same as Fig.~\protect\ref{R0_SIGMA_FIG}, except we plot
$r_1 \sqrt{\sigma}$.  Physically, the difference is that this
is sensitive to shorter distances than the previous plot.
}
\end{figure}

As a quantitative measure of the change in the shape of the potential due to the
dynamical quarks, we may plot dimensionless quantities such as $r_0 \sqrt{\sigma}$
or $r_1 \sqrt{\sigma}$.  Figures~\ref{R0_SIGMA_FIG} and \ref{R1_SIGMA_FIG}
show these two quantities as a function of quark mass, where $(m_\pi/m_\rho)^2$
is used as an indicator of the quark mass.  The effect of the sea quarks is
clear, and it should be noted that the effect is appreciable even for fairly
heavy sea quarks.  (The dependence of these quantities on the sea quark mass
emphasizes that only one dimensionful quantity can be matched in tuning full
and quenched simulations to match the lattice spacing.)
Figure~\ref{R0_SIGMA_FIG} can be compared with Fig.~1 of Ref.~\cite{SESAM_POT}, which
plots the same quantity with two flavors of Wilson quarks.
We note in passing that an extrapolation of Fig.~\ref{R0_SIGMA_FIG} to light quark
mass is consistent with commonly used values
for $r_0$ and $\sqrt{\sigma}$: $\ 0.50\ {\rm fm} \times 440\ {\rm MeV} = 1.12$.

\widetext
\narrowtext

\widetext
\narrowtext

\section{phenomenological consequences}

The larger quark-antiquark force at small distances in full QCD has been
suggested as a qualitative explanation for some of the differences between
quenched and full QCD that have been found in simulations.  For example,
there is some indication that dynamical quarks increase the value of
$f_B$ and $f_{B_S}$\cite{MILC_FB_ORIG,MILC_FB_PRL,NRQCD_FB,CPPACS_FB}, 
which can
be understood as a larger wave function at the origin due to the deeper
short distance potential well in full QCD\cite{FB_EFFECT}.
Also, dynamical quarks are expected to increase the spin splittings
in heavy quarkonia\cite{PSI_EFFECT,ONOGI_QUARKONIUM,CPPACS_QUARKONIUM}, 
which can again
be considered to result from a smaller wave function.

To make a crude estimate of these effects, we have solved the nonrelativistic
Schroedinger equation for the ground state in the static potentials.  Of course,
we don't advocate this as a replacement for a real solution to QCD, but in the
spirit of trying to understand the differences between quenched and full QCD,
it is a worthwhile exercise.  Specifically, we used the fitted Coulomb plus linear plus
constant form for the potential.
Alternative approaches, where we connected the points in the potential by straight
lines or solved a discrete lattice version of the Schroedinger equation, gave similar
results.
We show the ``Coulomb plus linear'' results because we expect that this is
closer to continuum physics than the discrete lattice potential, which is finite
even at $r=0$.  However, it must be remembered that this $1/r$ term is
phenomenological, not fundamental.  (We are working at distances too large
to identify it with a perturbative potential, and too short to identify it with a
universal correction to the linear potential.)
For light valence quark masses the wave function extends beyond the region where we
have fit the potential and into the region where we would expect string
breaking to show up.  By using the fitted form of the potential, we are ignoring
this effect.  Conversely, for heavy valence quark masses the wave function is small,
and sensitive to the treatment of the short distance potential.  However, with our
matched quenched and full QCD lattices, we expect this effect to be the same for
the quenched and full potentials.  Note that we can use our small sample of quenched
lattices with $a \approx 0.1$ fm to test the sensitivity to the lattice artifacts.

To repeat, a comparison of a dimensionful quantity between
quenched and full QCD can only be made subject to a choice of the quantity used
to determine the length scale in the quenched theory.  For example, if we choose
the string tension, then we might compare the wave functions at the origin in
units of the string tension, or $\Psi(0) \sigma^{-3/4}$.

\begin{figure}[tbp]
\epsfxsize=6.5in
\epsfysize=6.5in
\epsfbox[0 0 4096 4096]{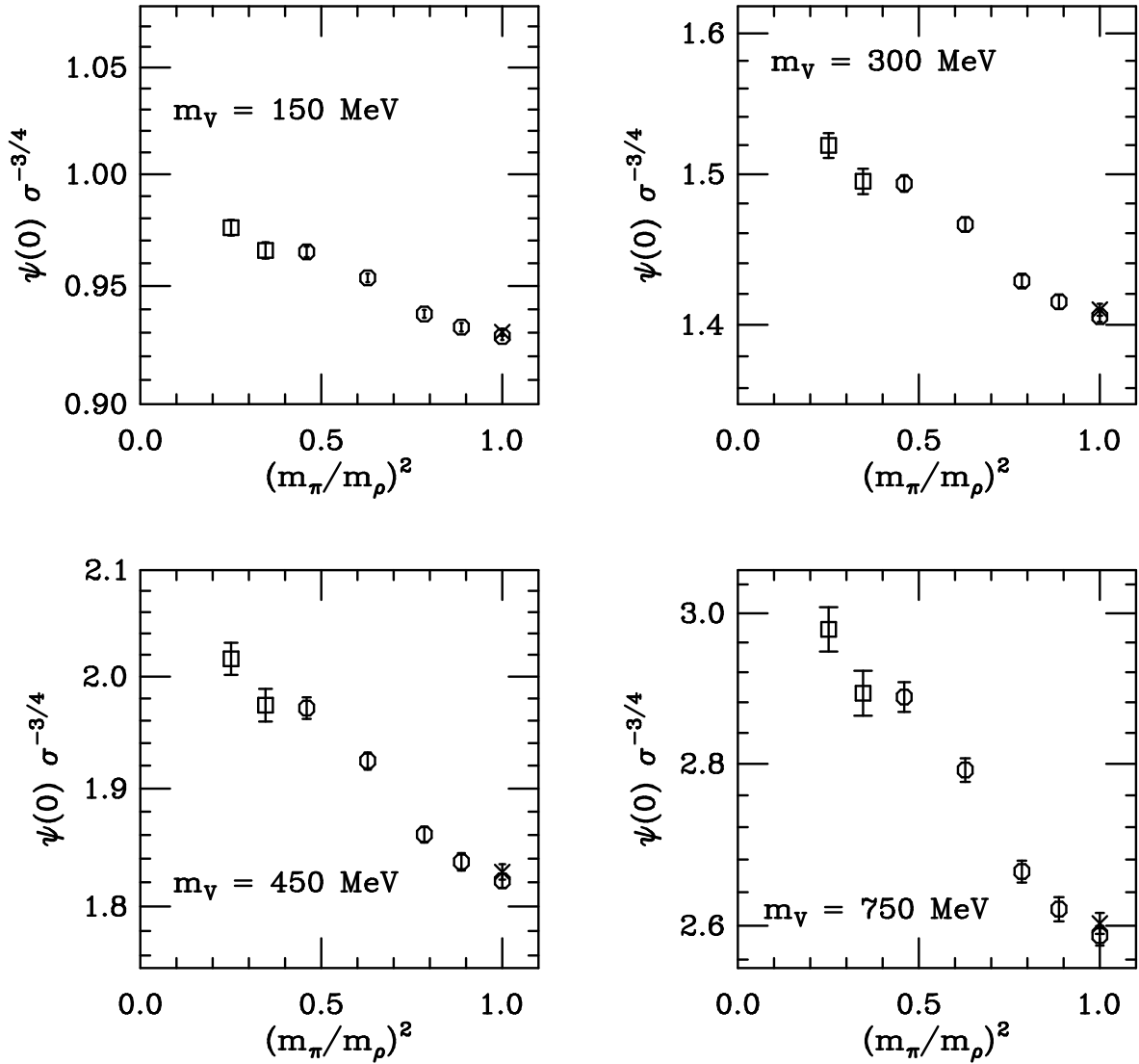}
\caption{
\label{PSI_0_FIG}
Wave function at the origin in units of the string tension for valence
quark masses of 150, 300, 450 and 750 MeV.
The horizontal scale is the sea quark mass, parameterized by $\LP m_\pi / m_\rho \RP^2$, 
placing the quenched calculation at the right side.
For the two smallest sea quark masses (squares), there are two light flavors of sea quark and
one heavier flavor at about the strange quark mass.  The cross indicates the result for
the quenched sample with $a \approx 0.10$ fm.
The vertical scale for each graph covers a range of twenty percent.
}
\end{figure}

We solved Schroedinger's equation for valence quark masses of 150, 300,
450 and 750 MeV.  These can be regarded as constituent quark masses
for, respectively, a light-light meson reduced mass (this is a stretch), a
heavy-light meson, a heavy-strange meson, and charmonium reduced mass.
Figure~\ref{PSI_0_FIG} shows the ground state wave function at the origin
in units of the string tension for these four masses as a function of
the sea quark mass, parameterized by $\LP m_\pi / m_\rho \RP^2$.
As expected, the larger force at short distance results in a larger wave
function at the origin in full QCD.  This is consistent with indications that
the inclusion of dynamical quarks increases decay constants
($f_B$ and $f_{B_S})$\cite{MILC_FB_ORIG,MILC_FB_PRL,NRQCD_FB,CPPACS_FB}
and hyperfine splittings in 
quarkonia\cite{PSI_EFFECT,ONOGI_QUARKONIUM,CPPACS_QUARKONIUM}.
This figure also shows $\Psi(0)$ from the quenched $a \approx 0.10$ fm run.
As expected, the effect of decreasing the lattice spacing is larger for
the heavier quark masses.
As can be seen from Fig.~\ref{R0_SIGMA_FIG}, Fig.~\ref{PSI_0_FIG} would look quite
different if we chose $r_0$ or $r_1$ to set the scale.
For example, at $m_q=150$ MeV. the wave function in units of $r_0$ increases
by less than one percent as the sea quark mass is lowered to $0.02/a$,
while in units of $\sigma$ it increases by 
almost five percent.

\begin{figure}[tbp]
\epsfxsize=6.5in
\epsfysize=6.5in
\epsfbox[0 0 4096 4096]{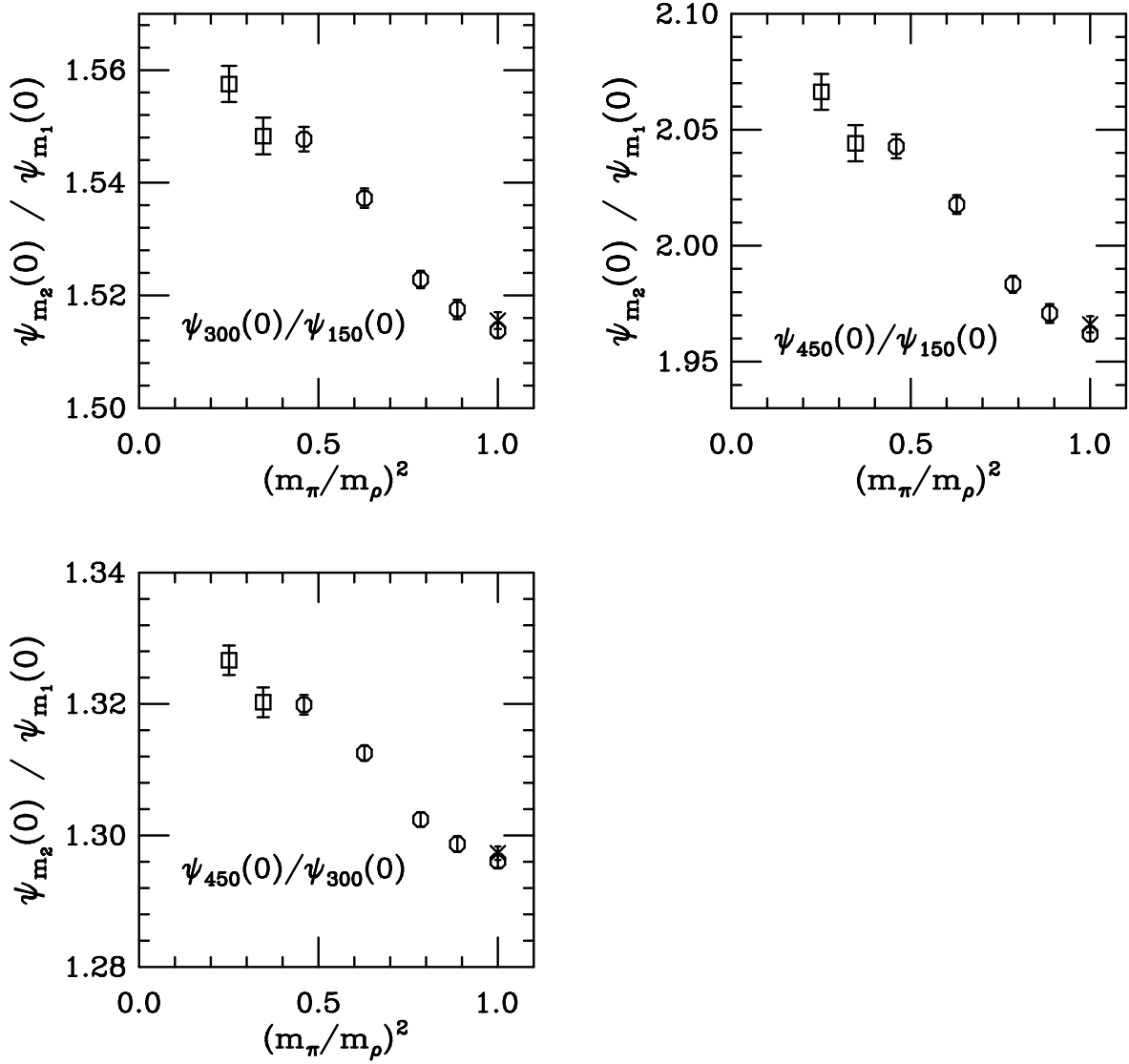}
\caption{
\label{PSI_RATIO_FIG}
Ratios of wave function at the origin for different valence quark masses.
The horizontal axis and symbols are the same as in Fig.~\protect\ref{PSI_0_FIG}.
The three panels show $\Psi_{300}(0)/\Psi_{150}(0)$, $\Psi_{450}(0)/\Psi_{150}(0)$
and $\Psi_{450}(0)/\Psi_{300}(0)$, respectively.  These can be {\bf very} loosely
interpreted as $f_B/f_\pi$, $f_{B_S}/f_\pi$ and $f_{B_S}/f_B$.
Note that in this figure the three panels cover different fractional ranges.
}
\end{figure}

An alternative approach is to use the wave function for light quarks
to set the scale.  This is similar to using $f_\pi$ to fix the lattice
spacing, as was done for example in the calculations of $f_B$ 
in Ref.~\cite{MILC_FB_PRL}.
Then we plot the ratios of wave functions for
different valence quark masses as functions of the sea quark masses
in Fig.~\ref{PSI_RATIO_FIG}.
Again, the dependence on sea quark mass is clearly visible, but the
size of the effect is much smaller than when the string tension sets
the scale.  For example, $\Psi(0) \sigma^{-3/4}$ at $m_V=300$ MeV increases by
$6.3\pm0.5\%$ with three degenerate flavors of sea quark with
$(m_\pi/m_\rho)^2=0.46$, or about the strange quark mass.  However,
the ratio of wave functions $\Psi_{300}(0)/\Psi_{150}(0)$ increases
by only $2.2 \pm 0.2\%$.
%FIXX: UPDATE THESE NUMBERS WHEN RUNS FINISH.

\widetext
\newpage
\newpage
\narrowtext

\widetext
\newpage
\newpage
\narrowtext

\section*{acknowledgements}

Computations were done on the Origin 2000 clusters at LANL,
on the T3E at NERSC, on the T3E, SP2 and SP3
at SDSC, on the NT cluster and Origin 2000 at NCSA, on the Origin
2000 at BU and on the PC cluster at Indiana University.
This work was supported by the U.S. Department of Energy under contracts
DOE -- DE-FG02-91ER-40628,	%C. Bernard
DOE -- DE-FG03-95ER-40894,	%T. DeGrand
DOE -- DE-FG02-91ER-40661,	%S. Gottlieb
DOE -- DE-FG05-96ER-40979 	%U. Heller
and
DOE -- DE-FG03-95ER-40906 	%D. Toussaint
and National Science Foundation grants
NSF -- PHY99-70701 		%C. DeTar
and
NSF -- PHY97--22022.    	%R. Sugar

\end{document}